\documentclass[english,twocolumn,aps,pra,superscriptaddress,address,showpacs,showacknowledgments]{revtex4-1}
\usepackage[T1]{fontenc}
\usepackage[latin9]{inputenc}
\usepackage{color}
\usepackage{amsmath}
\usepackage{amssymb}
\usepackage{graphicx}
\usepackage{babel}
\begin{document}

\title{Manifestation of classical nonlinear dynamics in optomechanical entanglement
with a parametric amplifier}

\author{Chang-Sheng Hu}

\affiliation{Fujian Key Laboratory of Quantum Information and Quantum Optics and
Department of Physics, Fuzhou University, Fuzhou 350116, People\textquoteright s
Republic of China}

\author{Li-Tuo Shen}

\affiliation{Fujian Key Laboratory of Quantum Information and Quantum Optics and
Department of Physics, Fuzhou University, Fuzhou 350116, People\textquoteright s
Republic of China}

\author{Zhen-Biao Yang}

\affiliation{Fujian Key Laboratory of Quantum Information and Quantum Optics and
Department of Physics, Fuzhou University, Fuzhou 350116, People\textquoteright s
Republic of China}

\author{Huaizhi Wu}

\affiliation{Fujian Key Laboratory of Quantum Information and Quantum Optics and
Department of Physics, Fuzhou University, Fuzhou 350116, People\textquoteright s
Republic of China}

\author{Yong Li}

\affiliation{Beijing Computational Science Research Center, Beijing 100193, People\textquoteright s
Republic of China}

\author{Shi-Biao Zheng} 

\affiliation{Fujian Key Laboratory of Quantum Information and Quantum Optics and
Department of Physics, Fuzhou University, Fuzhou 350116, People\textquoteright s
Republic of China}
\begin{abstract}
Cavity optomechanical system involving an optical parametric amplifier
(OPA) can exhibit rich classical and quantum dynamical behaviors.
By simply modulating the frequency of the laser pumping the OPA, we
find two interesting parameter regimes, with one of them enabling
to study quantum-classical correspondence in system dynamics, while
there exist no classical counterparts of the quantum features for
the other. For the detuning of the laser pumping around the mechanical
frequency, as the parametric gain of OPA increases to a critical value,
the classical dynamics of the optical or mechanical modes can experience
a transition from the regular periodic oscillation to period-doubling
motion, in which cases the light-mechanical entanglement can be well
studied by the logarithmic negativity and can manifest the dynamical
transition in the classical nonlinear dynamics. In addition, the optomechanical
entanglement shows a second-order transition characteristic at the
critical parametric gain. For the laser detuning being about twice
the mechanical frequency, the kind of normal mode splitting comes
up in the laser detuning dependence of optomechanical entanglement,
which is induced by the squeezing of the optical and mechanical hybrid
modes and finds no classical correspondence. The OPA assisted optomechanical
systems therefore offer a simple way to study and exploit quantum
manifestations of classical nonlinear dynamics.
\end{abstract}
\maketitle

\section{INTRODUCTION}

To seek for and explore quantum manifestations of classical nonlinear
dynamics is one of the most fundamental problems in physics. The familiar
dynamical behaviors, such as period-doubling, quasi-periodicity motions
and chaos \cite{Bak_prl2015,Ott_PT1994,Wang_prl2014}, which are very
common in classical nonlinear systems, can usually find their counterparts
in quantum systems. However, the quantum signature of classical nonlinear
dynamics may be found in different parameter regimes, leading to inconsistent
experimental requirements in studying classical and quantum dynamics.
For instance, some of the quantum chaotic problems, such as energy
level-spacing statistics \cite{Berr_Math1986} and quantum chaotic
scattering \cite{Lai_ibid1992}, can only be addressed in the limit
of low dimensional Hamiltonian \cite{Haa_b2010}. On the other hand,
quantum entanglement \cite{EPR_1935}, which is a fundamental phenomenon
in quantum mechanics, describes nonlocal two-body or many-body correlations
in a form of quantum superposition. However, the non-locality of quantum
entanglement has no classical correspondence. Much attention then
has been paid to the interplay between classical nonlinear dynamics
and quantum dynamical behaviors. Previous works have found that the
quantum-classical correspondence in the entanglement dynamics can
be analyzed from the perspective of classical trajectories and be
calculated by the reduced density linear entropy \cite{Zhang_pra2008},
and the quantum-phase transition in the Dicke model is accompanied
by the emergence of an entanglement singularity \cite{Lambert_PRL2004}.
In addition, the connections between classical collective dynamics
and quantum entanglement are widely studied in coupled qubit-resonator
systems \cite{Zhirov_prb2009}, trapped ions \cite{Lee_PRL2013},
and optomechanical systems \cite{Wang_prl2014}. 

Cavity optomechanics is associated with light-oscillator interacting
systems consisting of an optical cavity and a mechanical mirror \cite{As_Aspel_RMP2014,as_Mar_prl2007,as_Rae_prl2007,as_Teufel_Nature2011,as_Rabl_prl2011,as_Xu_pra2013,as_Li_oe2017}.
When the cavity is driven by an introduced input laser, the awakened
cavity will exert a radiation pressure on the movable mirror and make
it oscillate. Conversely, the oscillating mirror changes the length
of the cavity and thus the cavity intensity dependent radiation pressure
force, giving rise to the optomechanical coupling. Various interesting
quantum phenomena have been observed in optomechanical systems \cite{Blen_Phys2000,non_tan_pra2014,non_Liao_prl2016,non_Li_pra2018},
such as light-mechanical entanglement \cite{Vitali_prl2007,EN_Wnag_prl2018,EN_Chen_pra2014,EN_Liao_pra2015,EN_Li_prl2018,Macini_pra1997,Bose_pra1999,Zhang_pra2003,Pinard_EL2005}
and squeezed states \cite{La_Sci2004,sq_Kronwald_pra2013,sq_EE_Sci2015,sq_Lei_prl2016,Aga_pra2016,sq_Hu_pra2018}.
On the other hand, since the optomechanical coupling is intrinsically
nonlinear, the systems are good candidates for studying nonlinear
dynamical behaviors. Early experimental demonstration of chaotic dynamics
has been done by Carmon et al. \cite{Car_prl2007}. Before the dynamics
becomes chaotic, these optomechanical system may also experience some
times of period-doubling, going through the regular route from periodic
oscillations to quasi-periodicity, as discussed in some recent works
\cite{Larson_pra2011,Ma_pra2014,Lv_prl2015}. Moreover, a multi-mode
optomechanical system may exhibit collective dynamics, such as synchronization
and anti-phase synchronization \cite{Bem_pra2017,Mari_prl2013,Lud_prl2013,Liao_pra2019},
and as mentioned before, studies by using optomechanical systems as
a paradigm to address the correspondence between optomechanical entanglement
and synchronization have also been examined \cite{Ying_pra2014}.
In particular, strong fingerprints of periodic oscillations and quasi-periodic
motion in the quantum entanglement can be found \cite{Wang_prl2014}.

To search for the correspondence between entanglement dynamics and
classical nonlinear motion, we consider a composite cavity optomechanical
setting, in which an optical parametric amplifier (OPA) is involved.
The OPA, a second-order optical crystal in nature, has many applications
in both classical and quantum optics. In the classical case, it is
used to transform laser light into almost any (optical) frequency.
From a quantum point of view, the pairs of down-converted photons
generated by OPA can show nearly perfect single- or dual-squeezing
\cite{Gerry_book2005}. Therefore, the interplay between optomechanics
and OPA is naturally a significant topic to study the nonlinear quantum
dynamics. The OPA can modify the dynamical instabilities and nonlinear
dynamics of the system \cite{Pina_pra2016,Huang_pra2009_spll}. Moreover,
it has also found applications in generating strong mechanical squeezing
\cite{Aga_pra2016,sq_Hu_pra2018}, enhancing optomechanical cooling
\cite{Huang_pra2009}, and increasing single-photon optomechanical
coupling \cite{LV_prl2015_114}. 

We envision a two-field driving scenario, with one of them driving
the cavity and the other pumping the OPA. By scanning the laser frequency
of the OPA pumping, we find two interesting regimes. First, we find
good quantum-classical correspondence between the light-mechanical
entanglement and the classical nonlinear dynamics for the laser detuning
of OPA around the mechanical frequency. In this regime, when the classical
behavior transits from the regular periodic oscillation to period-doubling
motion as the parametric gain of OPA increases, the light-mechanical
entanglement dynamics also exhibits regular periodic oscillation and
period-doubling motion, in correspondence to the respective classical
behavior. In particular, we find a critical point $\Lambda_{c}$ for
the transition from periodic motion to period-doubling motion, which
is related to an entanglement singularity, namely, the light-mechanical
entanglement continuously enhances with the increase of the OPA gain,
but its derivative with respect to the parametric gain is discontinuous.
This characteristic is in resemblance to a second-order quantum phase
transition. Second, while the laser pumping of OPA is detuned from
cavity resonance by twice the mechanical frequency, the light-mechanical
entanglement exhibits normal mode splitting kind of feature, which
can be explained via the hybrid mode squeezing principle. Our result
gives a nice example of the manifestation of classical nonlinear dynamics
in optomechanical entanglement by simply modulating the OPA pumping,
and can be easily tested by the state-of-the-art cavity optomechanical
systems.

The rest of this paper is organized as follows. We first introduce
the optomechanical setup in Sec. II and then discuss the classical
dynamics in Sec. III. In Sec. IV, we introduce the measure of the
optomechanical entanglement and discuss the interesting physics with
the laser pumping of OPA around twice the mechanical frequency. In
Sec. V, we discuss both the quantum and classical dynamical behavior
for the laser pumping of OPA around the mechanical frequency, and
show the evidence for quantum-classical correspondence. We finally
summarize our result in Sec. VI.

\section{MODEL}

\begin{figure}
\includegraphics[width=1\columnwidth]{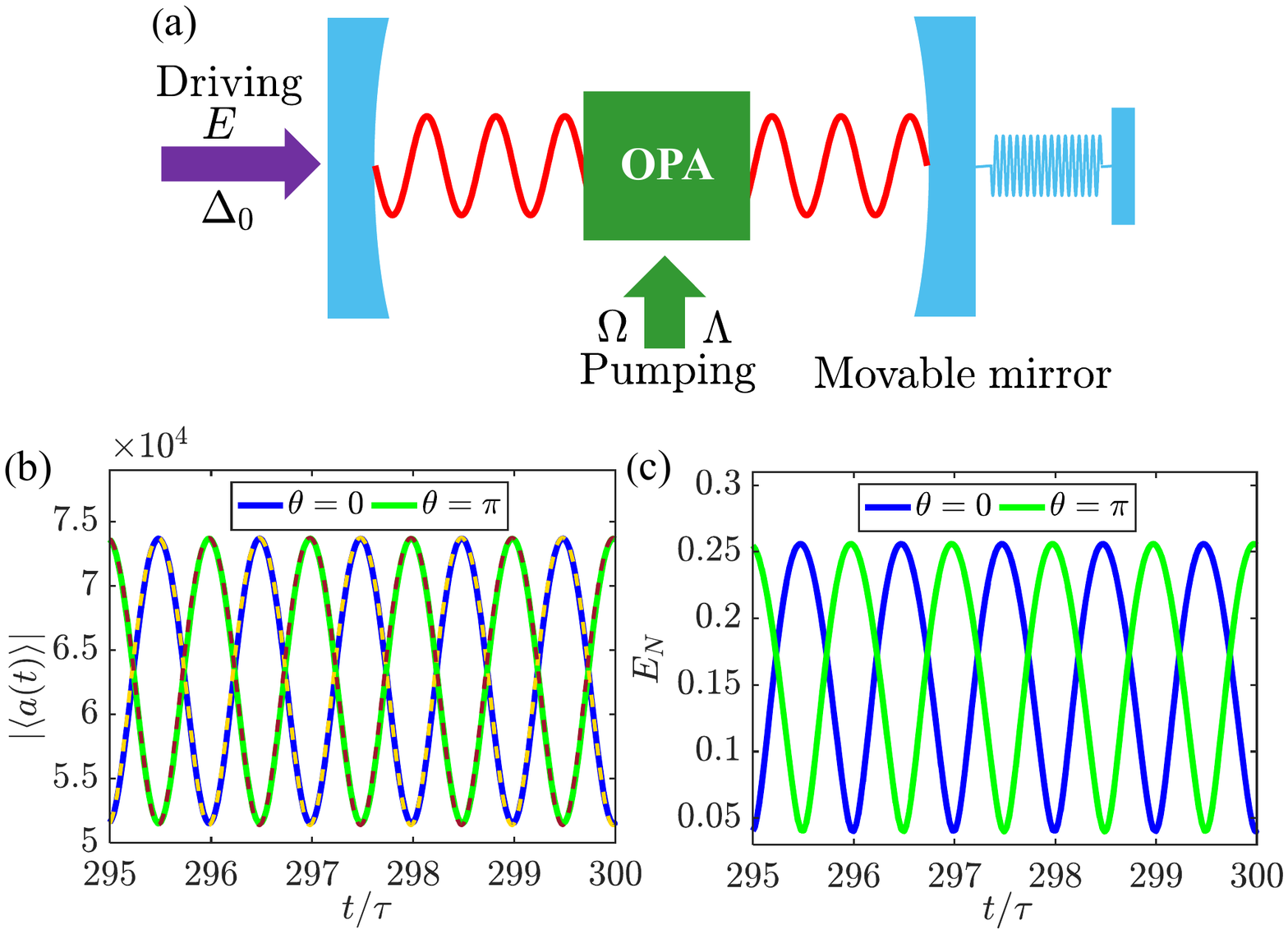}

\caption{\label{fig:Model}(Color online) (a) Sketch of the optomechanical
system under two-field driving. The cavity is driven by an external
driving laser with strength $E$ and detuning $\Delta_{0}$, and the
OPA inside the cavity is pumped by another laser field. (b) Long-time
behavior of the cavity field strength $|\langle a(t)\rangle|$ and
(c) the light-mechanical entanglement $E_{N}$ with $\tau=2\pi/\Omega$
for a typical set of experimental parameters \cite{Tho_Nat2008}:
$(E,g,\gamma_{m},\kappa,\Delta_{0})/\omega_{m}=(6\times10^{4},4\times10^{-6},10^{-6},0.2,1)$.
The OPA is pumped with the parametric gain $\Lambda/\omega_{m}=0.02$,
the detuning $\Omega/\omega_{m}=1.16$, and the parametric phase $\theta=0,\text{ \ensuremath{\pi}}$.
Note that the mechanical thermal noise is not taken into account so
far (i.e. $n_{m}=0$), however, it can be detrimental for experimental
generation of the optomechanical entanglement, see the Sec. V for
further discussions. For current consideration, the analytical approximation
of the classical mean values (dash) shows nice agreement with the
numerical solution. }
\end{figure}

We consider an optomechanical system with a degenerate optical parametric
amplifier (OPA) inside the Fabry-Perot cavity with one fixed partially
transmitting mirror and one movable totally reflecting mirror, as
shown in Fig. \ref{fig:Model}. Supposing that the length of the cavity
is $L$ in the rest and the finesse is represented by $F$, which
leads to a photon decay rate given by $\kappa=\pi c/(2FL)$. The movable
mirror is treated as a quantum-mechanical harmonic oscillator with
effective mass $m$, frequency $\omega_{m}$, and energy decay rate
$\gamma_{m}$. We assume that the cavity mode with the resonant frequency
$\omega_{c}$ is driven by an external laser of the frequency $\omega_{l}$
and the amplitude $E=\sqrt{2\kappa P/\hbar\omega_{l}}$, depending
on the laser power $P$. On the other hand, the degenerate OPA is
pumped by another laser field at frequency $2\omega_{p}$, giving
rise to a parametric gain denoted by $\Lambda e^{i\theta}$, which
is determined by the strength and the phase of the pumping laser.
The total Hamiltonian of the system in the rotating frame at the laser
frequency $\omega_{l}$ is given by 

\begin{eqnarray}
H & = & \hbar\Delta_{0}a^{\dagger}a+\frac{\hbar\omega_{m}}{2}(p^{2}+q^{2})-\hbar ga^{\dagger}aq+i\hbar E(a^{\dagger}-a)\nonumber \\
 &  & +i\hbar\Lambda(e^{i\theta}a^{\dagger2}e^{-i\Omega t}-e^{-i\theta}a^{2}e^{i\Omega t}).\label{eq:initial_Hamilton}
\end{eqnarray}
Here, $\Delta_{0}=\omega_{c}-\omega_{l}$ and $\Omega/2=\omega_{p}-\omega_{l}$
are the detunings of the driving laser frequency from the cavity resonance
and the down-conversion photon frequency; $a$ and $a^{\dagger}$
are the annihilation and creation operators of the cavity mode, respectively;
$q$ and $p$ are the position and momentum operators for the movable
mirror, satisfying the standard canonical commutation relation $[q,\:p]=i\hbar$;
and $g$=$x_{ZPF}\omega_{c}/L$ is the single-photon optomechanical
coupling strength arisen from the radiation pressure force with $x_{ZPF}=\sqrt{\frac{\hbar}{2m\omega_{m}}}$
being the zero point motion of the mechanical mode. In Eq. (\ref{eq:initial_Hamilton}),
the first and second terms are the free energies of the cavity mode
and the mechanical oscillator, respectively; the third term describes
the radiation-pressure induced coupling between the cavity and the
mechanical mode; the fourth term describes the longitudinal cavity
driving, and the last term represents the OPA pumping, leading to
the generation of pairs of cavity photons.

Using the Heisenberg equations of motion for the cavity and mechanical
operators and taking the mechanical damping and cavity decay into
account, the dynamics of the system can be described by the following
set of quantum Langevin equations \cite{Bar_book2004} 

\begin{eqnarray}
\dot{q} & = & \omega_{m}p,\nonumber \\
\dot{p} & = & -\omega_{m}q-\gamma_{m}p+ga^{\dagger}a+\xi(t),\nonumber \\
\dot{a} & = & -(\kappa+i\Delta_{0})a+igaq+E+2\Lambda e^{i\theta}a^{\dagger}e^{-i\Omega t}\nonumber \\
 &  & +\sqrt{2\kappa}a_{in}(t),\label{eq:Langevin}
\end{eqnarray}
where $a_{in}$ is the zero-mean vacuum-input noise operator satisfying
the auto-correlation function \cite{Bar_book2004} $\langle a_{in}^{\dagger}(t)a_{in}(t')\rangle=\delta(t-t')$,
and ${\textstyle \xi}(t)$ is the thermal noise operator acting on
the mechanical oscillator. In the regime of high mechanical quality
factor $Q\equiv\omega_{m}/\gamma_{m}\gg1$, the Markovian approximation
can be applied to the thermal noise, leading to \cite{Giovan_pra2001}
$\left\langle \xi(t)\xi(t')+\xi(t')\xi(t)\right\rangle /2=\gamma_{m}(2n_{m}+1)\delta(t-t')$
with $n_{m}=[\exp(\hbar\omega_{m}/k_{B}T)-1]^{-1}$ being the mean
thermal excitation number in the mechanical mode and $T$ the thermal
bath temperature.\textcolor{black}{{} Note that, for $\Omega\neq0$,
the parametric interaction here plays a role in resemblance to a periodically
amplitude-modulated cavity driving \cite{Mari_prl2009}, but the OPA-modulated
amplitude depends on the time-varying cavity field itself via the
term $\sim2\Lambda a^{\dagger}$. I}n the following sections, we focus
on the connections between typical dynamical behavior of the classical
mean values and the light-mechanical entanglement when the parametric
gain and the driving frequency of OPA are especially tuned.

\section{Classical steady-state behavior }

\begin{figure}
\includegraphics[width=1\columnwidth]{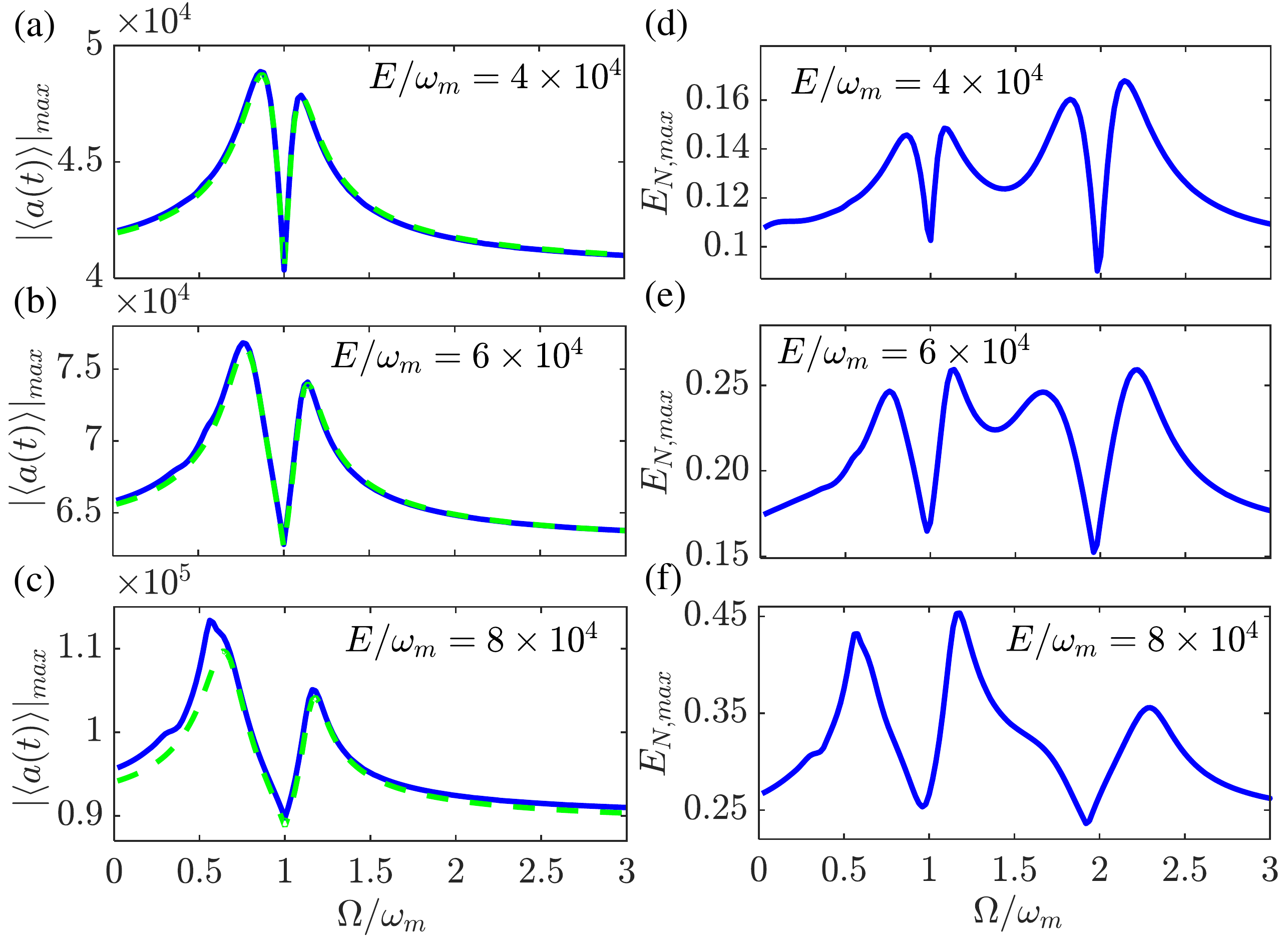}\caption{\label{fig:cav_EN_Ome}(Color online) $|\langle a(t)\rangle|_{max}$
{[}(a)-(c){]} and $E_{N,max}$ {[}(d)-(f){]} in the long time limit
as a function of $\Omega/\omega_{m}$ for driving strengths $E/\omega_{m}=4\times10^{4}$
{[}(a),(d){]}, $E/\omega_{m}=6\times10^{4}$ {[}(b),(e){]}, and $E/\omega_{m}=8\times10^{4}$
{[}(c),(f){]}. The numerical (analytical) results are presented by
the blue solid (green dashed) lines. Other parameters are the same
as in Figs. \ref{fig:Model}(b), \ref{fig:Model}(c).}
\end{figure}

We next derive the steady-state solution $\left\langle O(t)\right\rangle $
$(O=q,\,p,\,a)$ for the Eq. (\ref{eq:Langevin}) via the mean-field
theory, where $\left\langle a(t)q(t)\right\rangle =\left\langle a(t)\right\rangle \left\langle q(t)\right\rangle $
and $\left\langle a^{\dagger}(t)a(t)\right\rangle =\left\langle a^{\dagger}(t)\right\rangle \left\langle a(t)\right\rangle $.
The equations of motion Eq. (\ref{eq:Langevin}), which are in analogy
to the amplitude-modulated driving schemes \cite{sq_Hu_pra2018,Fara_pra2012,Mari_prl2009,Mari_njp2012},
may allow the classical mean values $\left\langle O(t)\right\rangle $
to evolve toward an asymptotic periodic orbit with the periodicity
$\tau=2\pi/\Omega$, which can then be analyzed via the Fourier series
expansion if the parameters keep the system far from optomechanical
instabilities and multistabilities. Since the effectively modulated
amplitude $2\Lambda|\langle a^{\dagger}(t)\rangle|\propto2\Lambda E$
corresponding to OPA pumping ($\sim e^{-i\Omega t}$) in Eq. (\ref{eq:Langevin})
is much weaker than the driving field $E$, we then derive the steady-state
solutions of Eq. (\ref{eq:Langevin}) in terms of \textcolor{black}{the
power series ansatz in }$\varepsilon_{s}=2\Lambda E$ as $\left\langle O(t)\right\rangle =\sum_{j=0}^{\infty}\sum_{n=-\infty}^{\infty}O_{n,j}\varepsilon_{s}^{j}e^{in\Omega t}$,
with $O_{n,j}$ being time independent coefficients \cite{sq_Hu_pra2018,Fara_pra2012,Mari_prl2009,Mari_njp2012}.
For weak cavity drivings, we can further rewrite $\left\langle O(t)\right\rangle $
to the first order in $\varepsilon_{s}$ and set $n=0,\pm1$, leading
to \cite{Huang_pra2010}
\begin{eqnarray}
\left\langle O(t)\right\rangle  & = & O_{0}+\varepsilon_{s}e^{-i\Omega t}O_{+}+\varepsilon_{s}^{*}e^{i\Omega t}O_{-}.\label{eq:Fourier}
\end{eqnarray}
Substituting Eq. (\ref{eq:Fourier}) into the mean-field version of
Eq. (\ref{eq:Langevin}), it is readily to get the time-independent
coefficients $O_{0}$, $O_{+}$, and $O_{-}$
\begin{equation}
p_{0}=0,\quad q_{0}=\frac{g|a_{0}|^{2}}{\omega_{m}},\quad a_{0}=\frac{E}{\kappa+i\Delta},\label{eq:asymp0}
\end{equation}

\begin{eqnarray}
a_{+} & = & \frac{1}{d(\Omega)\left(\kappa-i\Delta\right)}\{[\kappa-i(\Delta+\Omega)](\omega_{m}^{2}-\Omega^{2}-i\gamma_{m}\Omega)\nonumber \\
 &  & +ig^{2}\omega_{m}|a_{0}|^{2}\}e^{i\theta},\label{eq:asymp1}
\end{eqnarray}
\begin{equation}
a_{-}=\frac{ig^{2}\omega_{m}a_{0}^{2}e^{-i\theta}}{d^{*}(\Omega)\left(\kappa+i\Delta\right)},\label{eq:asymp-1}
\end{equation}
where $\Delta=\Delta_{0}-gq_{0}$ is the effective detuning and 
\begin{eqnarray}
d(\Omega) & = & -2\Delta g^{2}\omega_{m}|a_{0}|^{2}+[\kappa-i(\Delta+\Omega)][\kappa+i(\Delta-\Omega)]\nonumber \\
 &  & \times(\omega_{m}^{2}-\Omega^{2}-i\gamma_{m}\Omega).\label{eq:d_Ome}
\end{eqnarray}
Note that the expressions for $q_{\pm}$ and $p_{\pm}$ are not explicitly
given since they are unrelated to our discussion in the following.
Using the analytical approximations for the steady-state mean values,
we show in Fig. \ref{fig:Model}(b) the comparison of the time-evolutional
dynamics with the asymptotic solutions and the full numerical results,
which agree well with each other in the weak nonlinear regime. Without loss of generality, the phase of the pumping laser is set to zero in other parts of the paper.

\begin{figure}
\includegraphics[width=1\columnwidth]{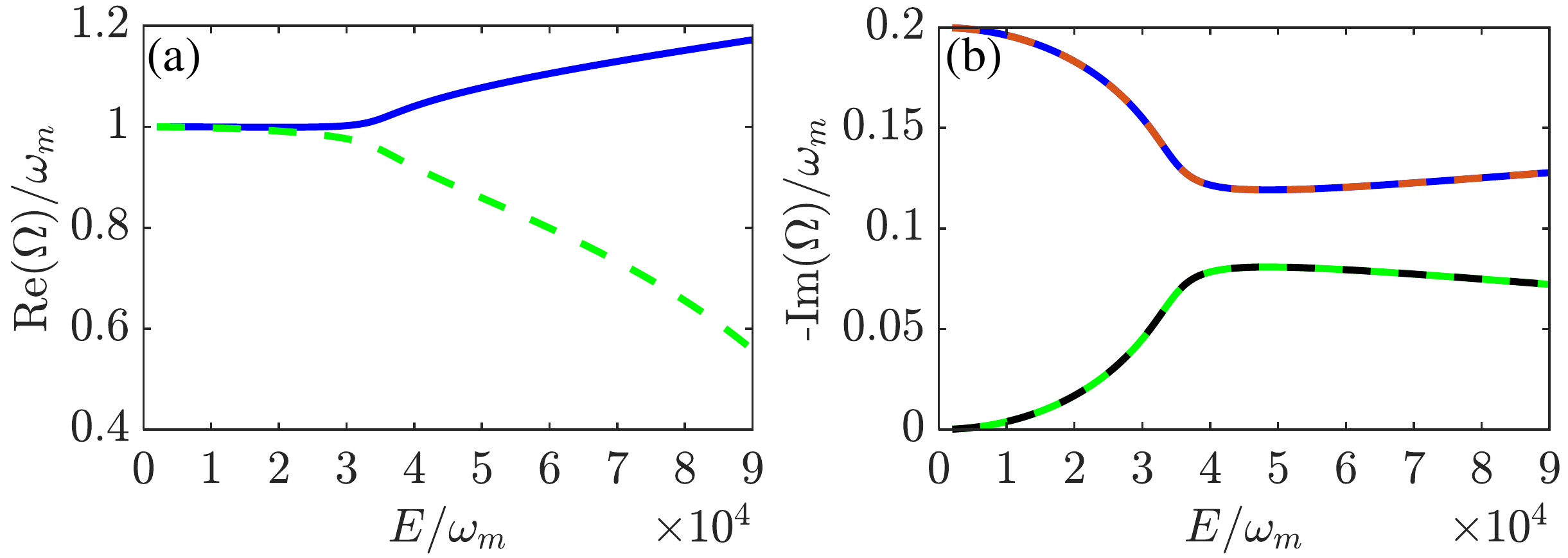}\caption{\label{fig:root}(Color online) (a) The real parts of roots of $d(\Omega)=0$
in the domain $\text{Re}(\Omega)>0$, and (b) the imaginary parts
of the roots of $d(\Omega)=0$ as a function of driving $E/\omega_{m}$.
Other parameters are the same as in Fig. \ref{fig:cav_EN_Ome}.}
\end{figure}

To examine the effect of strong nonlinear dynamics, we plot $|\langle a(t)\rangle|_{max}=\underset{\tau}{max}\{|\langle a(t)\rangle|\}$,
which is the maximum of $|\langle a(t)\rangle|$ in one period $\tau$
in the long time limit, as a function of the pumping detuning $\Omega$
in Figs. \ref{fig:cav_EN_Ome}(a)-(c) for gradually increased driving
strengths $E$. We find a normal-mode-splitting-like feature, where
the separation of the two peaks increases as the driving power enhances.
The splitting of $|\langle a(t)\rangle|_{max}$ that arises as a result
of the optomechanical coupling $g$ is determined by the structure
of the denominator in Eq. (\ref{eq:d_Ome}). Thus, the peaks' position
can be evidenced by the real parts of the roots of $d(\Omega)=0$
in the domain $\text{Re}(\Omega)>0$, as shown in Fig. \ref{fig:root}(a).
When the driving strength $E$ is small, the real parts of the roots
of $d(\Omega)=0$ have two equal values, so no splitting appears.
However, there is a splitting in the imaginary part of the roots,
i.e. the lifetime splitting \cite{Gup_OC1995}, see Fig. \ref{fig:root}(b).
Increasing the driving strength $E$ to a certain value, the real
parts of the roots in the domain $\text{Re}(\Omega)>0$ begin to split.
The difference of the two values monotonically increases with the
driving strength $E$ being enhanced. 

\textcolor{black}{In general, the asymptotic solutions Eqs. (\ref{eq:asymp0})-(\ref{eq:asymp-1})
can be good approximations of the system steady-state behavior for
a wide range of laser detuning $\Omega$ for OPA pumping with $E/\omega_{m}=4\times10^{4},\text{ }6\times10^{4}$.
While for $E/\omega_{m}=8\times10^{4}$, the higher-order terms $\sim\varepsilon_{s}^{j}e^{in\Omega t}$
(with $j>1$ or $n>1$) in the Fourier expansion solution should be
included for a better fittig. Thus, the first-order approximation
Eq. (\ref{eq:Fourier}) can only provide the proof-of-principle study
of the classical nonlinear dynamics. Furthermore, when the cavity
intensity $|\langle a(t)\rangle|^{2}$ exceeds the threshold that
leads to the instability of mechanical motion, the nonlinear system
dynamics cannot be regarded as the quasi-periodic behavior \cite{Mari_prl2009,Mari_njp2012}
and the Fourier expansion solution of the steady-state values $\left\langle O(t)\right\rangle $
will be invalid. The studies in the following thus focus on the parameter
regime, where the long-time dynamics of the system is asymptotically
periodic and the optomechanical instability is avoided, which will
be numerically studied by checking the Routh-Hurwitz criterion while
the system's temporal evolution approaches the dynamical steady state,
see further discussions later.}

\section{Optomechanical entanglement}

By linearizing the dynamics around the classical mean values of the
operators, i.e. $O(t)=\langle O(t)\rangle+\delta O(t)$, we can find
the equations of motion for the quantum fluctuations, which are given
by
\begin{eqnarray}
\delta\dot{q} & = & \omega_{m}\delta p,\nonumber \\
\delta\dot{p} & = & -\omega_{m}\delta q-\gamma_{m}\delta p+\frac{1}{\sqrt{2}}[G(t)\delta a^{\dagger}+G(t)^{*}\delta a]+\xi(t),\nonumber \\
\delta\dot{a} & = & -[\kappa+i\Delta(t)]\delta a+\frac{i}{\sqrt{2}}G(t)\delta q+2\Lambda e^{i\theta}\delta a^{\dagger}e^{-i\Omega t}\nonumber \\
 &  & +\sqrt{2\kappa}a_{in}(t),\label{eq:Linearized Langevin}
\end{eqnarray}
where $G(t)=\sqrt{2}g\left\langle a(t)\right\rangle $ is the photon-number-dependent
optomechanical coupling, and $\Delta(t)=\Delta_{0}-g\left\langle q(t)\right\rangle $
is the effective detuning of the cavity resonance from the driving
laser frequency. Owing to all of the quantum information is included
in Eq. (\ref{eq:Linearized Langevin}) when the system is in the stability
regime, one can quantitatively measure the light-mechanical entanglement
by the linearized equations. Distinguished from the standard optomechanical
system \cite{Wang_prl2014,Vitali_prl2007}, the OPA can enrich the
dynamics of quantum entanglement and allow us to characterize the
quantum manifestation of the classical dynamics. 

We measure the degree of light-mechanical entanglement by the logarithmic
negativity $E_{N}$. For our purposes, we use a vector $u(t)=[\delta q,\delta p,\delta x,\delta y]^{T}$
including the variables of the quantum fluctuations, in which $\delta x=(\delta a+\delta a^{\dagger})/\sqrt{2}$,
$\delta y=(\delta a-\delta a^{\dagger})/i\sqrt{2}$ represent the
amplitude and phase quadratures of the cavity mode, respectively,
with the corresponding noise quadratures being $\delta x_{in}=(\delta a_{in}+\delta a_{in}^{\dagger})/\sqrt{2}$,
$\delta y_{in}=(\delta a_{in}-\delta a_{in}^{\dagger})/i\sqrt{2}$.
Then the time-dependent equations of motion for the quantum fluctuations
can be expressed as $\dot{u}(t)=M(t)u(t)+n(t),$ with the drift matrix
\begin{equation}
M(t)=\left[\begin{array}{cccc}
0 & \omega_{m} & 0 & 0\\
-\omega_{m} & -\gamma_{m} & \text{Re}G(t) & \text{Im}G(t)\\
-\text{Im}G(t) & 0 & -\kappa+R_{c}(t) & \Delta(t)-R_{s}(t)\\
\text{Re}G(t) & 0 & -\Delta(t)-R_{s}(t) & -\kappa-R_{c}(t)
\end{array}\right],
\end{equation}
where $R_{c}(t)=2\Lambda\cos(\Omega t-\theta)$, $R_{s}(t)=2\Lambda\sin(\Omega t-\theta)$
and the input noise $n(t)=[0,\xi(t),\sqrt{2\kappa}\delta x_{in},\sqrt{2\kappa}\delta y_{in}]^{T}$.
Without parametric interaction (i.e. $\Lambda=0$), the drift matrix
$M$ is time independent, the formal solution for $u(t)$ is simply
given by $u(t)=e^{Mt}u(0)+\int_{0}^{t}dt^{\prime}e^{Mt^{\prime}}n(t-t^{\prime})$,
and the system is stable and reaches its stationary steady state for
$t\rightarrow\infty$ when all the eigenvalues of $M$ have negative
real parts. For the time-dependent drift matrix here (i.e. $\Lambda\neq0$
and $\Omega\neq0$), the sufficient condition for stability then corresponds
to the fact that all eigenvalues of $M(t)$ have negative real parts
for the system approaching the long-time periodic dynamics, which
associates with the Routh-Hurwitz criterion. While the system is in
stable, the optomechanical interaction strength $G(t)$, the effective
couplings $-R_{s}\pm\Delta$ and decays $-\kappa\pm R_{c}$ for the
cavity quadratures $\delta x,\text{ \ensuremath{\delta}}y$ are then
modulated periodically by the OPA pu\textcolor{black}{mping, leading
to strong fingerprints in the second moments of the quadratures of
the classical nonlinear dynamical behaviors.}

Due to the zero-mean Gaussian nature of the quantum noises, the equations
of motion for $u(t)$ can further be reformulated in terms of the
$4\times4$ covariance matrix $V(t)$ with $V_{k,l}(t)=\langle u_{k}(t)u_{l}^{\dagger}(t)+u_{l}^{\dagger}(t)u_{k}(t)\rangle/2$
and the diagonal noise correlations matrix $D=\text{diag}[0,\gamma_{m}(2n_{m}+1),\kappa,\kappa]$
\cite{Mari_prl2009}, leading to $\dot{V}(t)=M(t)V(t)+V(t)M^{T}(t)+D$.
To calculate $E_{N}$, the covariance matrix is expressed as a $2\times2$
sub-block matrices $V(t)=[V_{1},V_{c};V_{c}^{T},V_{2}]$, where $V_{1}$,
$V_{2}$, and $V_{c}$ correspond to the mechanical mode, cavity mode,
and the optomechanical correlation, respectively \cite{Wang_prl2014}.
Finally, the logarithmic negativity $E_{N}$ can be calculated by
\cite{Ying_pra2014}
\begin{eqnarray}
E_{N} & \equiv & \textrm{max}[0,\,-\text{ln}2\eta],\label{eq:Measure_EN}
\end{eqnarray}
with $\eta\equiv(1/\sqrt{2})[\Sigma(V)-\sqrt{\Sigma(V)^{2}-4\text{det}V}]^{1/2}$
and $\Sigma(V)=\text{det}(V_{1})+\text{det}(V_{2})-2\text{det}(V_{c})$.
Since the drift matrix $M(t)$ strongly depends on the parameters
$\Lambda,\text{ }\Omega$ of the OPA pumping, the OPA-modulated driven-dissipative
dynamics can therefore strongly modify the nonlinear behavior of the
optomechanical entanglement. In what follows, we first assume the
thermal phonon number to be zero (i.e. $n_{m}=0$), nevertheless,
it should be noted that the thermal noise is one of the main detrimental
effects to optomechanical entanglement generation in experiments,
which will be studied later by considering a finite thermal temperature
(see Sec. V).

The temporal dynamics of the light-mechanical entanglement $E_{N}$
is shown in Fig. \ref{fig:Model}(c), where the time-periodic $E_{N}$
oscillates with the periodicity $\tau=2\pi/\Omega$ in the stable
regime, resembling to the mean-field dynamics of $|\langle a(t)\rangle|$.
We then quantify the light-mechanical entanglement $E_{N}$ by the
maximum in one period $\tau$ \cite{Wang_prl2014,Fara_pra2012,Mari_prl2009,Mari_njp2012}
namely
\begin{eqnarray}
E_{N,max} & = & \underset{\tau}{max}\{E_{N}(t)\}.\label{eq:E_N_max}
\end{eqnarray}
Similarly, we plot $E_{N,max}$ as a function of the detuning $\Omega$
in Figs. \ref{fig:cav_EN_Ome}(d)-(f) for different driving strengths
$E$. By comparing them with the classical result $|\langle a(t)\rangle|_{max}$,
one can observe two peaks around $\Omega/\omega_{m}=2$, which are
classically nonexistent. To see the insight, we rewrite the effective
optomechanical coupling as $G(t)=g_{0}+g_{+1}e^{-i\Omega t}+g_{-1}e^{i\Omega t}$
by using Eq. (\ref{eq:Fourier}), and assume $g_{0}\propto a_{0}$,
$g_{\pm1}\propto2\Lambda Ea_{\pm}$ to be positive reals without loss
of generality. In fact, the third term $\sim g_{-1}$ can be further
neglected when $g_{-1}\ll g_{0},\text{ }g_{+1},\Omega$. Using Eq.
(\ref{eq:Linearized Langevin}), and with $\delta q=(\delta b+\delta b^{\dagger})/\sqrt{2}$
and $\delta p=(\delta b-\delta b^{\dagger})/i\sqrt{2}$, the effective
linearized Hamiltonian for $\Delta\approx\omega_{m}$ after neglecting
the fast oscillating terms becomes $H_{eff}=(-1/2)[g_{0}\delta a\delta b^{\dagger}+g_{+1}e^{i(\Omega-2\omega_{m})t}\delta a\delta b+i2\Lambda e^{-i\theta}e^{i(\Omega-2\omega_{m})t}\delta a^{2}]+H.c$.
We then introduce the normal modes $c_{\pm}=(\delta a\pm\delta b)/\sqrt{2}$
and rewrite the Hamiltonian in the interaction picture by rotating
with $\left(-g_{0}/2\right)(c_{+}^{\dagger}c_{+}-c_{-}^{\dagger}c_{-})$.
Finally, we arrive at
\begin{eqnarray}
H_{eff} & \approx & \frac{1}{4}g_{\pm}c_{\pm}^{2}+H.c.,\label{eq:2modeSqueezing}
\end{eqnarray}
for $\Omega-2\omega_{m}\pm g_{0}=0$, where $g_{\pm}=-(g_{+1}\pm2i\Lambda e^{-i\theta}$).\textcolor{black}{{}
Eq. (\ref{eq:2modeSqueezing}) which associates with the $squeezing$
of the two hybrid modes $c_{\pm}$ \cite{Mari_njp2012}, gives rise
to the two peaks around $\Omega/\omega_{m}=2$ in the $\Omega$-dependent
light-mechanical entanglement, as shown in Fig. \ref{fig:cav_EN_Ome}
(d)-(f). Owing to $g_{0}\propto E$, the separation of the two peaks
increases with the increasing of the driving strength $E$. }

\begin{figure}
\includegraphics[width=1\columnwidth]{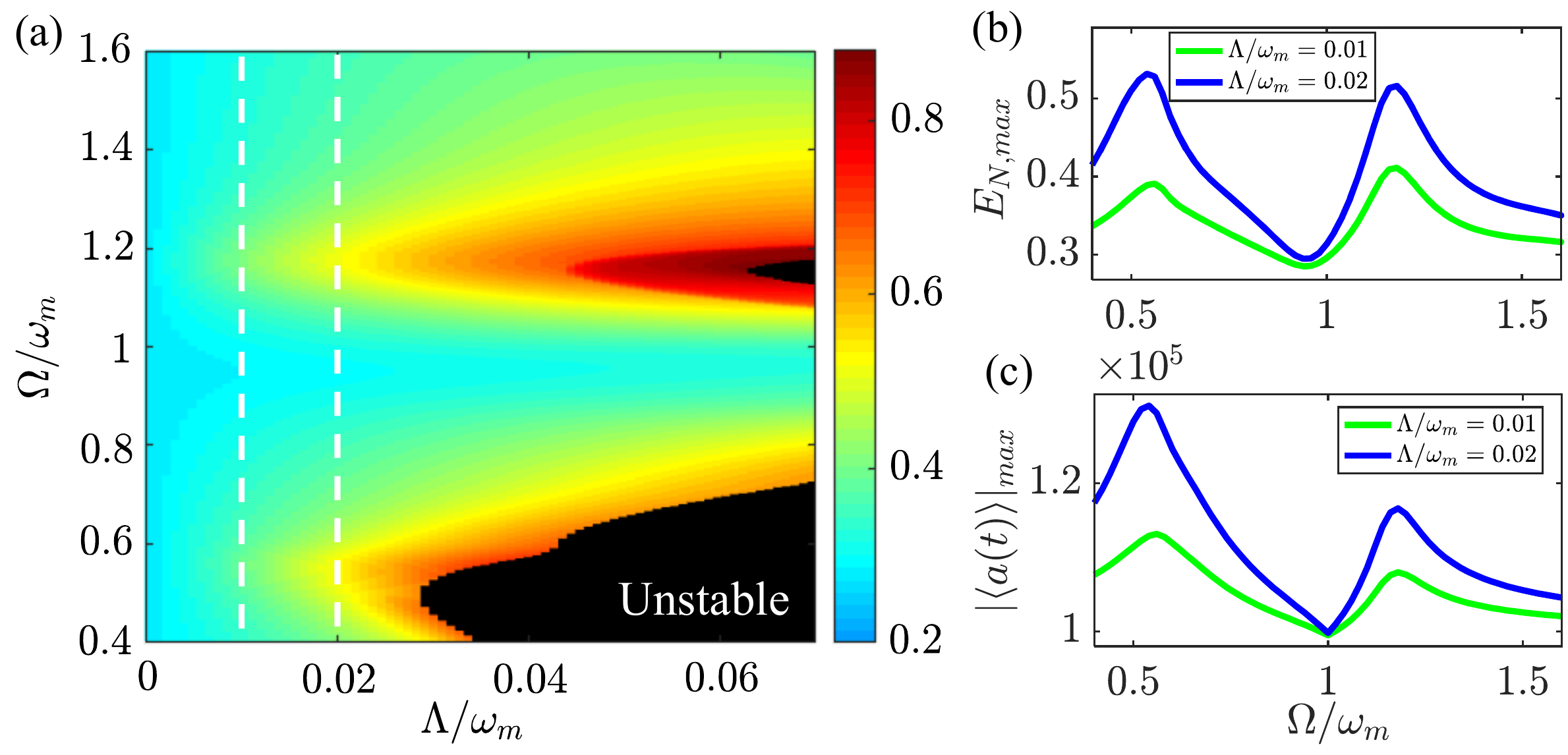}

\caption{\label{fig_3D}(Color online) (a) $E_{N,max}$ versus the parametric
gain $\Lambda/\omega_{m}$ and the detuning $\Omega/\omega_{m}$ for
$E/\omega_{m}=8.6\times10^{4}$. The black area denotes the classically
unstable region, confirmed \textcolor{black}{via the Routh-Hurwitz
criterion. (b) $E_{N,max}$ and (c) $|\langle a(t)\rangle|_{max}$
versus $\Omega/\omega_{m}$ for parametric gains $\Lambda/\omega_{m}=0.01$
(green) and $\Lambda/\omega_{m}=0.02$ (blue), as indicated by the
dashed lines in (a). Other} parameters are the same as in Figs. \ref{fig:Model}(b),
\ref{fig:Model}(c).}
\end{figure}

On the other hand, it is worthwhile to note that the light-mechanical
entanglement $E_{N}$ shows additional ridges near $\Omega/\omega_{m}=1$,
which correspond to the positions of the two peaks for the classical
mean value of cavity field strength $|\langle a(t)\rangle|_{max}$.
Therefore, it allows us to investigate the quantum manifestations
of classical nonlinear dynamics by selecting $\Omega$ around $\omega_{m}$.
To see stronger nonlinear effects, we will further increase the driving
strength $E$, however, the stability of the system will be checked
and maintained. In Fig. \ref{fig_3D}(a), we plot $E_{N,max}$ versus
the parametric gain $\Lambda/\omega_{m}$ and the detuning $\Omega/\omega_{m}$
for $E/\omega_{m}=8.6\times10^{4}$. In the stable regime, the peak
values of $E_{N}$ can be found at $\Omega/\omega_{m}\approx0.55$
and $\Omega/\omega_{m}\approx1.2$ for different $\Lambda/\omega_{m}$.
In addition, $E_{N,max}$ can be improved by increasing $\Lambda$,
as shown in Fig. \ref{fig_3D}(b) and the maximal $E_{N,max}$ is
realized in the vicinity of unstable regime with $\Omega/\omega_{m}\approx1.2$.
While for $\Omega/\omega_{m}\approx0.55$ the system quickly accesses
to the unstable regime with the increasing of $\Lambda$ and the entanglement
$E_{N,max}$ is much smaller. The classical correspondence is again
found in the laser-detuning $\Omega$ dependence of $|\langle a(t)\rangle|_{max}$,
see Fig. \ref{fig_3D}(c).

\begin{figure}
\includegraphics[width=1\columnwidth]{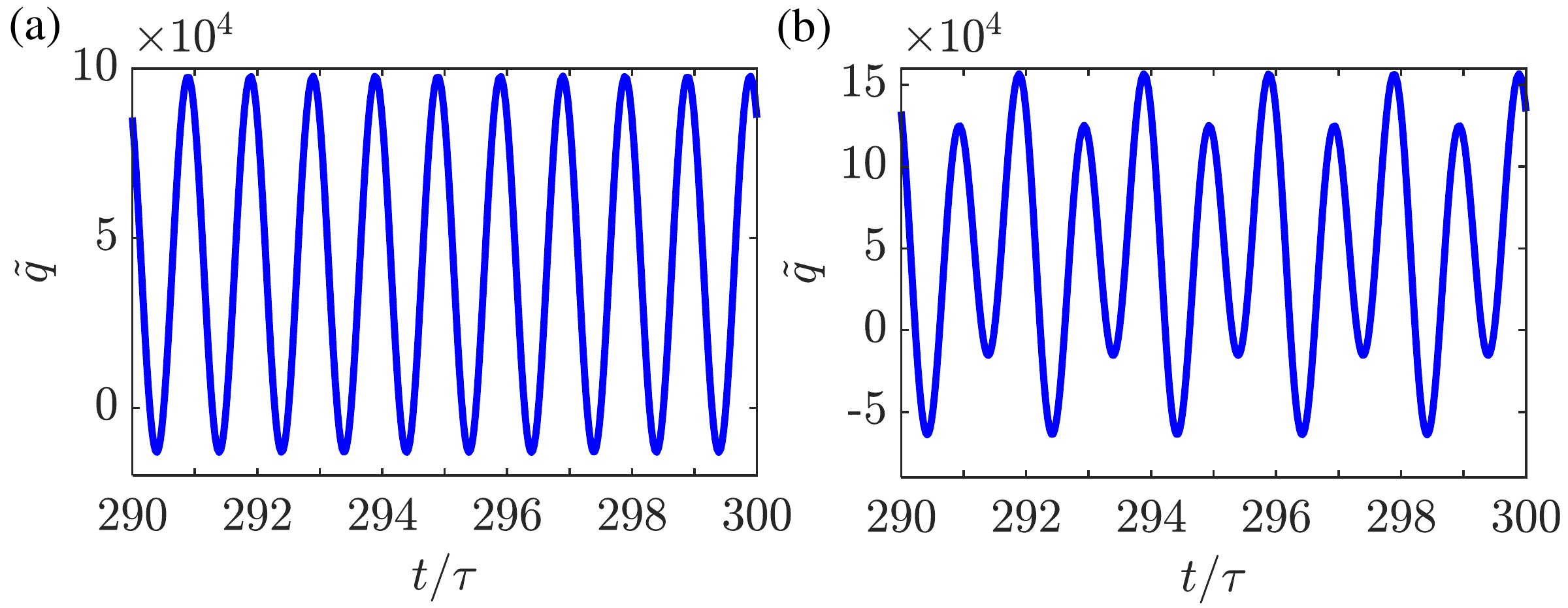}

\caption{\label{fig_two_q}(Color online) Long-time dynamics of the mechanical
coordinate $\tilde{q}$, which displays the regular periodic behavior
for $\Lambda/\omega_{m}=0.03$, and the period-doubling motion for
$\Lambda/\omega_{m}=0.05$ with the two-field detuning being $\Omega/\omega_{m}=1.16$
and the cavity driving strength being $E/\omega_{m}=8.6\times10^{4}$.
Other parameters are the same as in Figs. \ref{fig:Model}(b), \ref{fig:Model}(c).}
\end{figure}

\section{Quantum manifestations of classical nonlinear dynamics}

The classical dynamics discussed in Sec. III is based on the fact
that the asymptotic behaviors of the optical mode and mechanical motion
are approximately sinusoidal. However, it is known that the system
evolution will follow the route, namely, from periodic motion to chaotic
behavior when the laser pump of the OPA is gradually increasing. To
further study the classical nonlinear behavior, we now use the notations
$\tilde{p}$, $\tilde{q}$, and $\alpha=\alpha_{r}+i\alpha_{i}$ for
simplicity, which correspond to the mean values $\langle q(t)\rangle$,
$\langle p(t)\rangle$ and $\langle a(t)\rangle$, respectively, and
follow the equations of motion

\begin{equation}
\begin{array}{ccl}
\dot{\tilde{q}} & = & \omega_{m}\tilde{p},\\
\dot{\tilde{p}} & = & -\omega_{m}\tilde{q}-\gamma_{m}\tilde{p}+g(\alpha_{r}^{2}+\alpha_{i}^{2}),\\
\dot{\alpha}_{r} & = & (\Delta_{0}-g\tilde{q})\alpha_{i}-\kappa\alpha_{r}+E+2\Lambda\text{cos}(\Omega t-\theta)\alpha_{r}\\
 &  & -2\Lambda\text{sin}(\Omega t-\theta)\alpha_{i},\\
\dot{\alpha}_{i} & = & (-\Delta_{0}+g\tilde{q})\alpha_{r}-\kappa\alpha_{i}-2\Lambda\text{cos}(\Omega t-\theta)\alpha_{i}\\
 &  & -2\Lambda\text{sin}(\Omega t-\theta)\alpha_{r}.
\end{array}\label{eq:mean Langevin-1}
\end{equation}
The intensity of the cavity field is thus given by $I_{c}=\alpha_{r}^{2}+\alpha_{i}^{2}$,
and the dynamical behavior can be now characterized by the evolution
of the adjacent trajectories in phase space. For this purpose, let's
first introduce a new vector $\overrightarrow{\varepsilon}=(\varepsilon_{\tilde{q}},\varepsilon_{\tilde{p}},\varepsilon_{\alpha_{r}},\varepsilon_{\alpha_{i}})$
to describe how a tiny perturbation of the initial classical values
evolves in the phase space over time. The evolution equations of $\vec{\varepsilon}$
can be obtained by Eq. (\ref{eq:mean Langevin-1}) and are given by

\begin{equation}
\begin{array}{ccl}
\dot{\varepsilon}_{\tilde{q}} & = & \omega_{m}\varepsilon_{\tilde{p}},\\
\dot{\varepsilon}_{\tilde{p}} & = & -\omega_{m}\varepsilon_{\tilde{q}}-\gamma_{m}\varepsilon_{\tilde{p}}+2g(\alpha_{i}\varepsilon_{\alpha_{i}}+\alpha_{r}\varepsilon_{\alpha_{r}}),\\
\dot{\varepsilon}_{\alpha_{r}} & = & (\Delta_{0}-g\tilde{q})\varepsilon_{\alpha_{i}}-g\alpha_{i}\varepsilon_{\tilde{q}}-\kappa\varepsilon_{\alpha_{r}}+2\Lambda\text{cos}(\Omega t-\theta)\varepsilon_{\alpha_{r}}\\
 &  & -2\Lambda\text{sin}(\Omega t-\theta)\varepsilon_{\alpha_{i}},\\
\dot{\varepsilon}_{\alpha_{i}} & = & (-\Delta_{0}+g\tilde{q})\varepsilon_{\alpha_{r}}+g\alpha_{r}\varepsilon_{\tilde{q}}-\kappa\varepsilon_{\alpha_{i}}-2\Lambda\text{cos}(\Omega t-\theta)\varepsilon_{\alpha_{i}}\\
 &  & -2\Lambda\text{sin}(\Omega t-\theta)\varepsilon_{\alpha_{r}}.
\end{array}\label{eq:mean Langevin-1-1}
\end{equation}

By simulating Eq. (\ref{eq:mean Langevin-1}), we find that the temporal
evolution of the mechanical coordinate $\tilde{q}$ can transit from
the sinusoidal oscillation of the period $\tau=2\pi/\Omega$ to a
period-doubling oscillation for the laser driving $E/\omega_{m}=8.6\times10^{4}$
and appropriate OPA pumping strengths $\Lambda$. As shown in Fig.
\ref{fig_two_q}(a), the oscillatory period of $\tilde{q}$ is $\tau$
for $\Lambda/\omega_{m}=0.03$, while the evolution period becomes
twice $\tau$ for $\Lambda/\omega_{m}=0.05$ {[}see Fig. \ref{fig_two_q}(b){]}.
Note that the analytical solutions in Eq. (\ref{eq:Fourier}), which
built on the assumption of a sinusoidal oscillation, are not applicable
for the motion with the period $2\tau$.

\begin{figure}
\includegraphics[width=0.9\columnwidth]{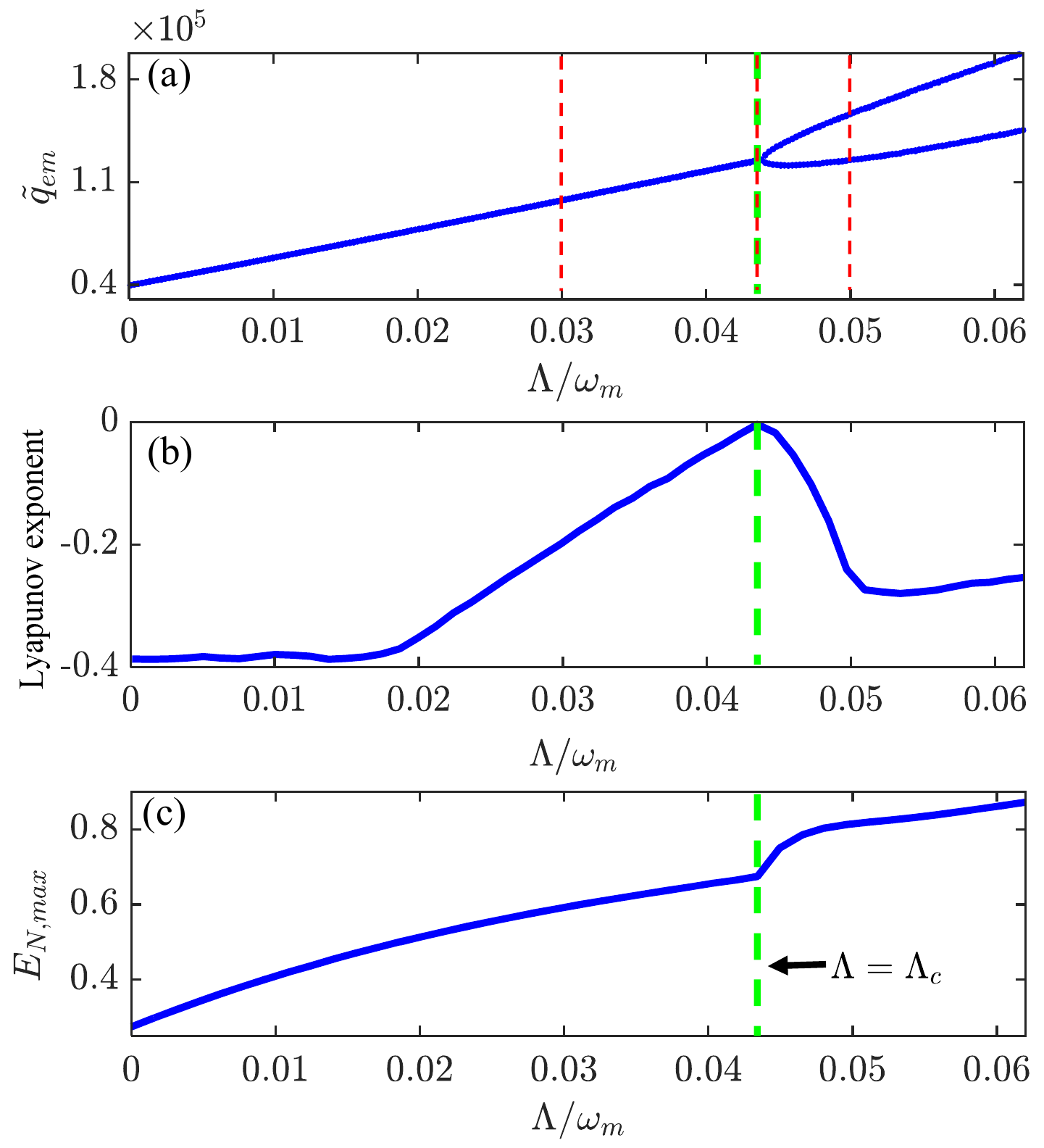}

\caption{\label{fig_bifurcation}(Color online) (a) Asymptotic extreme values
of the mechanical coordinate $\tilde{q}_{em}$ in the classical dynamics,
(b) Lyapunov exponent, and (c) $E_{N,max}$ versus parametric gain
$\Lambda/\omega_{m}$. Typical bifurcation diagram is found in $\tilde{q}_{em}$.
The green dash line indicates the critical point $\Lambda_{c}/\omega_{m}=0.043$
at which the transition from the regular periodic to period-doubling
motion occurs. The Lyapunov exponent is taken from the slope of the
time-evolutional curve of $\text{ln}(\varepsilon_{I_{c}})$ for the
time interval {[}$280\tau$, $300\tau${]} by using linear fitting,
with $\vec{\varepsilon}(t=0)=(10^{-10},10^{-10},10^{-10},10^{-10})$.
Other parameters are the same as in Fig. \ref{fig_two_q}. The red
dashed lines indicate the parametric regime discussed in Figs. \ref{fig_003}-\ref{fig_005}. }
\end{figure}

As a route to chaos, the period-doubling phenomena were previously
discussed in standard optomechanical systems \cite{Ma_pra2014} by
focusing on the classical nonlinear regime. Here, the main concern
is about the quantum entanglement near the classical transitions.
For chaotic dynamics, the light-mechanical entanglement can not be
measured by the logarithmic negativity any more since the Lyapunov
exponent is positive, see further discussion below. Thus, we devote
the task to the quantum manifestations of the two common types of
classical nonlinear behaviors: regular periodic oscillations and period-doubling
motion. However, the entanglement dynamics in chaotic regime may be
studied by other methods, such as trajectory-based calculation of
the density linear entropy \cite{Xu_pra2017}. In Fig. \ref{fig_bifurcation}(a),
we show the asymptotic extreme values $\tilde{q}_{em}$ of the dynamical
variable $\tilde{q}$ as a function of the parametric gain $\Lambda$
with Eq. (\ref{eq:mean Langevin-1}). Due to the OPA-modulated driven-dissipative
dynamics, the unique extreme value $\tilde{q}_{em}$ separates into
two branches as the parametric gain sweeps through the critical OPA
pumping $\Lambda_{c}$, and the oscillation amplitude continuously
increases for $\Lambda>\Lambda_{c}$.

Furthermore, we study the time-dependence of the Lyapunov exponent
\cite{Oseledec_TMM1968}, which is defined by the curve slope of the
natural logarithms of $\varepsilon_{I_{c}}(t)=\varepsilon_{\alpha_{r}}^{2}+\varepsilon_{\alpha_{i}}^{2}+2\alpha_{r}\varepsilon_{\alpha_{r}}+2\alpha_{i}\varepsilon_{\alpha_{i}}$
versus the evolutional time. The Lyapunov exponent shows how the initial
states of cavity field evolve in temporal domain and phase space {[}see
Fig. \ref{fig_bifurcation}(b){]} \cite{Ma_pra2014}. One can see
that the Lyapunov exponent is negative for $\Lambda<\Lambda_{c}$.
While the pumping strength approaches the transition point, the Lyapunov
exponent goes to zero. For $\Lambda>\Lambda_{c}$, where the period-doubling
motion is found, the Lyapunov exponent falls below zero again. In
general, there is no positive Lyapunov exponent for the pumping strength
under consideration, which implies that the Gaussian nature of the
random noise is well maintained for the quantum Langevin equation
and the quantum fluctuation (and therefore the quantum entanglement)
can be safely characterized by using the covariance matrix \cite{Wang_prl2014}.

\begin{figure}
\includegraphics[width=1\columnwidth]{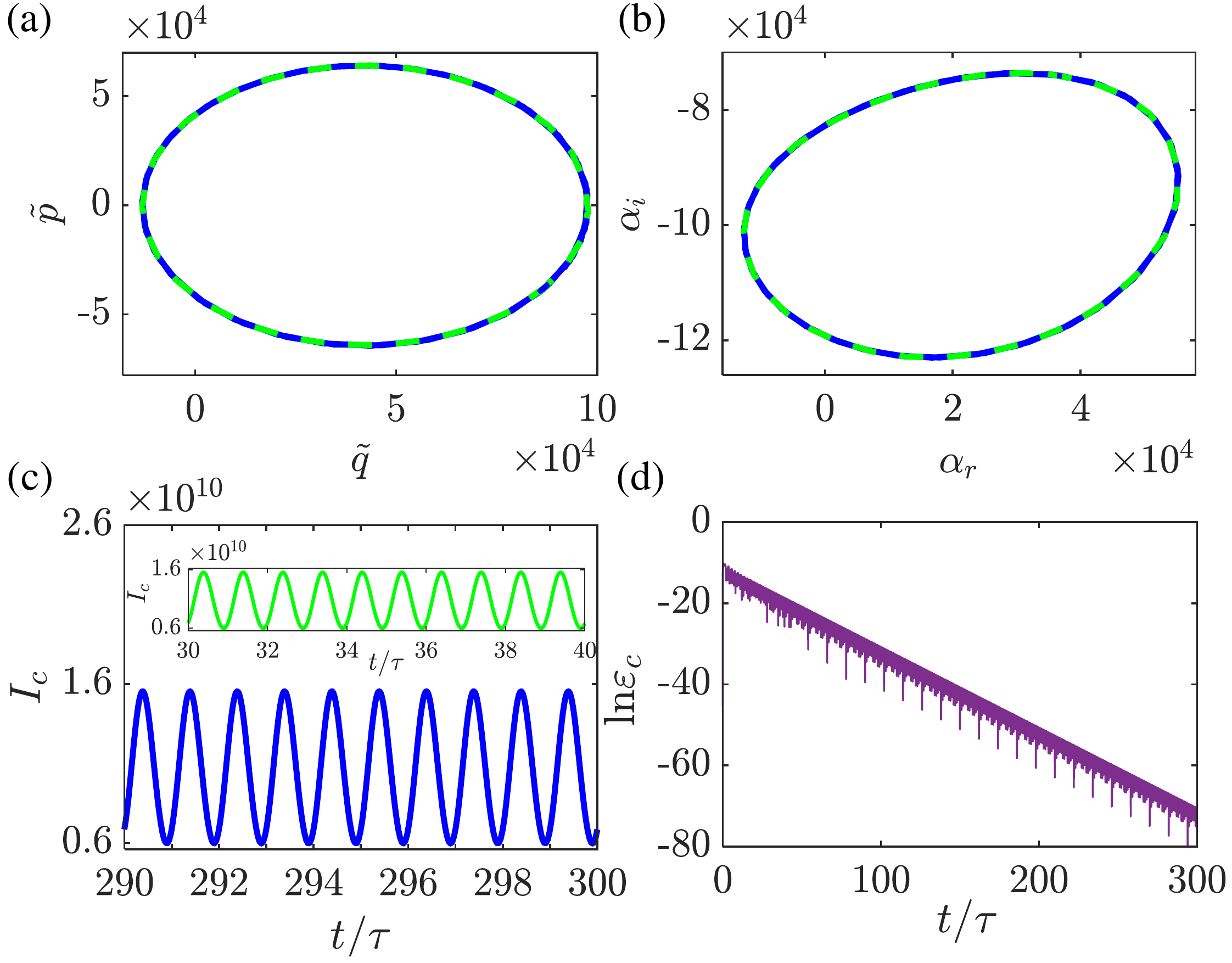}

\caption{\label{fig_003}(Color online) Phase space trajectories of (a) the
mechanical mode and (b) the cavity mode for time intervals {[}$30\tau$,
$40\tau${]} (green dashed lines), {[}$290\tau$,$300\tau${]} (blue
lines). (c) Time-dependence of the cavity intensity $I_{c}$ for the
time interval {[}$290\tau$, $300\tau${]} in comparison with that
for the interval {[}$30\tau$, $40\tau${]} (the inset). (d) Time-dependence
of $\text{ln}(\varepsilon_{I_{c}})$ in {[}$0$,$300\tau${]}. Parameters
are $\Lambda/\omega_{m}=0.03$, $\vec{\varepsilon}(t=0)=(10^{-10},10^{-10},10^{-10},10^{-10})$.
Other parameters are the same as in Fig. \ref{fig_two_q}.}
\end{figure}
\begin{figure}
\includegraphics[width=1\columnwidth]{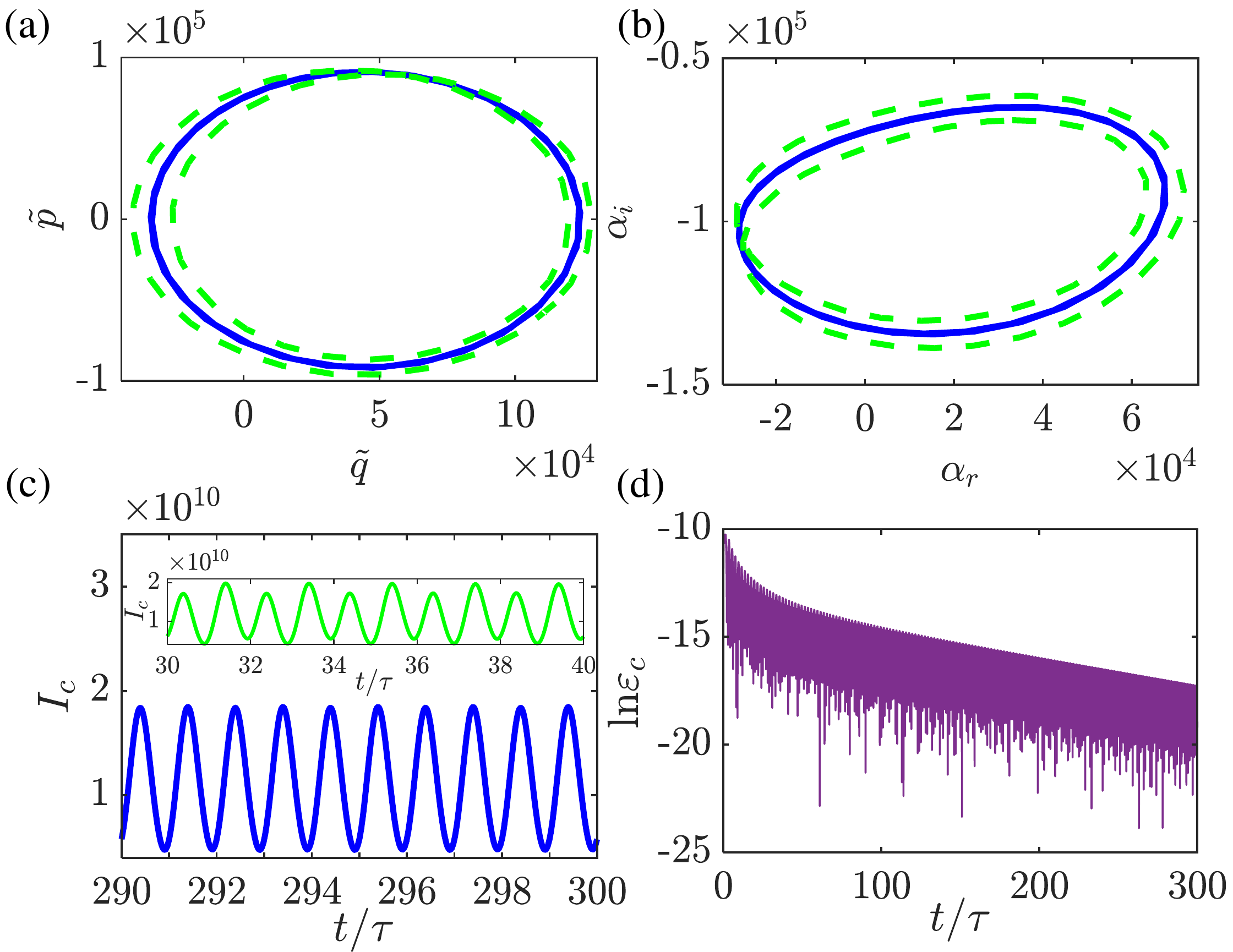}

\caption{\label{fig_0043}(Color online) Phase space trajectories of (a) the
mechanical mode and (b) the cavity mode for time intervals {[}$30\tau$,
$40\tau${]} (green dashed lines), {[}$290\tau$,$300\tau${]} (blue
lines). (c) Time-dependence of the cavity intensity $I_{c}$ for the
time interval {[}$290\tau$, $300\tau${]} in comparison with that
for the interval {[}$30\tau$, $40\tau${]} (the inset). (d) Time-dependence
of $\text{ln}(\varepsilon_{I_{c}})$ in {[}$0$,$300\tau${]}. Parameters
are $\Lambda/\omega_{m}=0.043$, $\vec{\varepsilon}(t=0)=(10^{-10},10^{-10},10^{-10},10^{-10})$.
Other parameters are the same as in Fig. \ref{fig_two_q}.}
\end{figure}
\begin{figure}
\includegraphics[width=1\columnwidth]{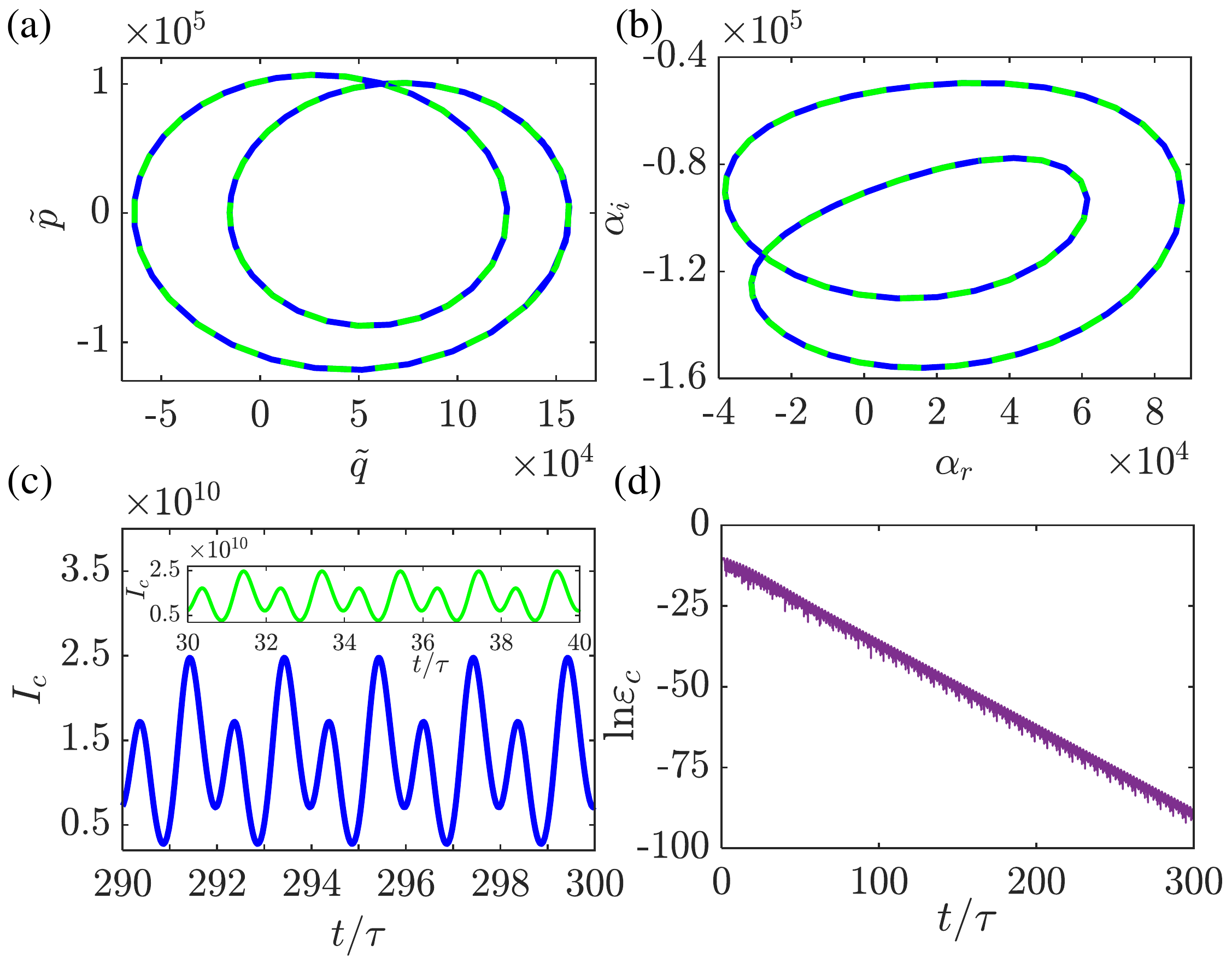}

\caption{\label{fig_005}(Color online) Phase space trajectories of (a) the
mechanical mode and (b) the cavity mode for time intervals {[}$30\tau$,
$40\tau${]} (green dashed lines), {[}$290\tau$,$300\tau${]} (blue
lines). (c) Time-dependence of the cavity intensity $I_{c}$ for the
time interval {[}$290\tau$, $300\tau${]} in comparison with that
for the interval {[}$30\tau$, $40\tau${]} (the inset). (d) Time-dependence
of $\text{ln}(\varepsilon_{I_{c}})$ in {[}$0$,$300\tau${]}, respectively.
Parameters are $\Lambda/\omega_{m}=0.05$, $\vec{\varepsilon}(t=0)=(10^{-10},10^{-10},10^{-10},10^{-10})$.
Other parameters are the same as in Fig. \ref{fig_two_q}.}
\end{figure}

Fig. \ref{fig_bifurcation}(c) shows the maximum light-mechanical
entanglement $E_{N,max}$ as a function of $\Lambda$. One can see
that before transition the quantum entanglement smoothly increases
with the increasing parametric gain $\Lambda$, but a ``cusp'' appears
at the transition point $\Lambda_{c}$, signified by a sudden increase
of entanglement. After the transition, the classical dynamics becomes
period-doubling motion, and the quantum entanglement recovers the
smoothly increasing characteristic. The general feature resembles
to the classical bifurcation followed by the satu\textcolor{black}{ration
to a stable attractor induced by nonlinear interactions \cite{Wang_prl2014}.
However, this is described as a characteristic of second-order phase
transition in quantum mechanics since the derivatives of the quantum
entanglement versus $\Lambda$ are not continuous near the transition
point \cite{Ying_pra2014}. This implies that a second-order, cusp
type of phase transition in light-mechanical entanglement occurs by
the aid of OPA in this optomechanical system. As a result, the OPA-modulated
driven-dissipative dynamics evokes a classically nonlinear period-doubling
behavior accompanied by the appearance of a second order transition
in optomechanical entanglement at the critical OPA gain. }

\begin{figure}
\includegraphics[width=1\columnwidth]{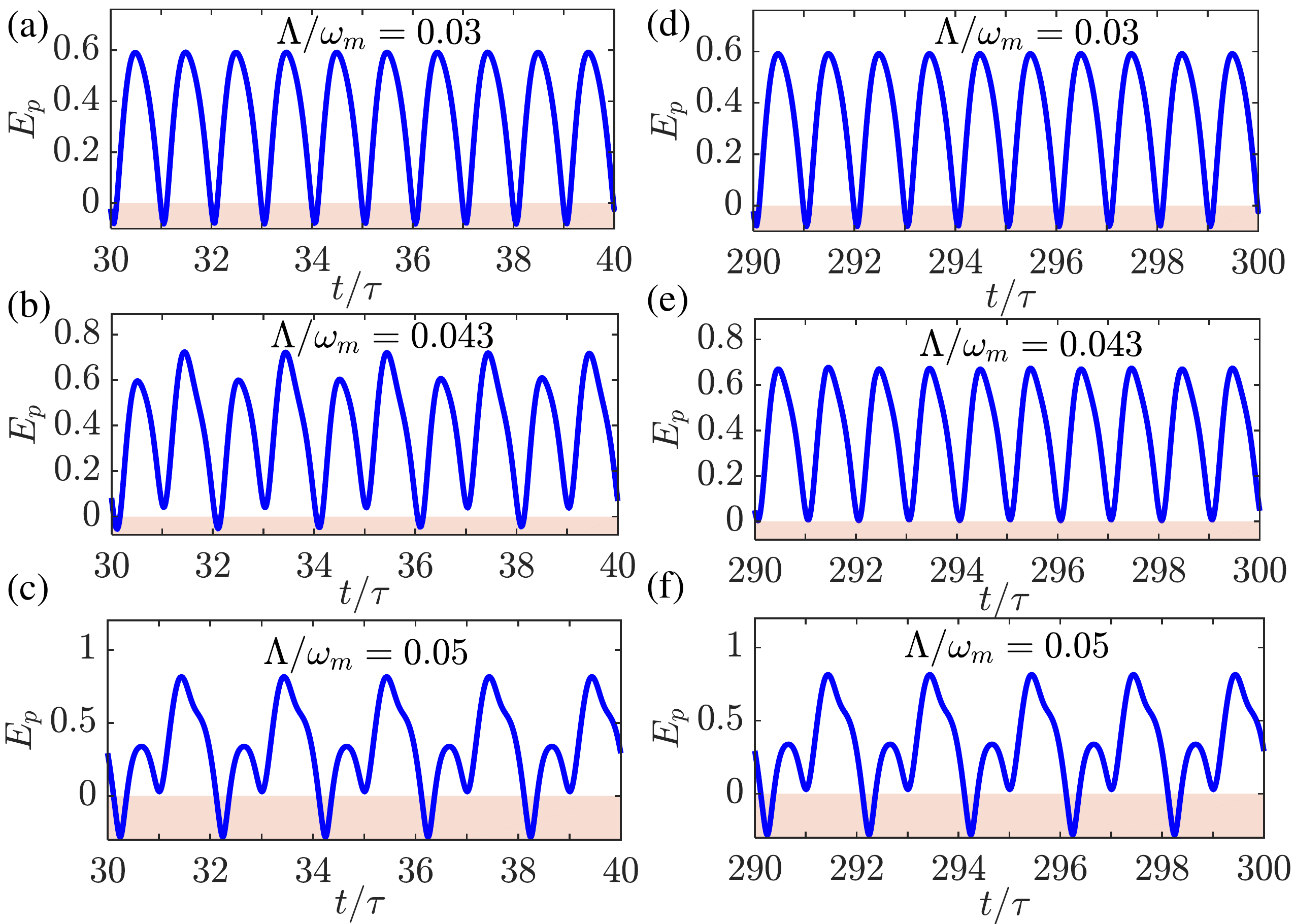}

\caption{\label{EN_three}(Color online) Short-time dynamics {[}(a)-(c){]}
and long-time dynamics {[}(d)-(f){]} of $E_{P}$ for $\Lambda/\omega_{m}=0.03$
{[}(a),(d){]}, $0.043$ {[}(b),(e){]}, and $0.05$ {[}(c),(f){]}.
Other parameters are the same as in Fig. \ref{fig_bifurcation}.}
\end{figure}

To further investigate the dynamical behavior near the transition
point in more detail, we show in Figs. \ref{fig_003}-\ref{fig_005}
the steady-state phase-space trajectory of the mechanical and cavity
modes, the steady-state cavity intensity $I_{c}$, and the logarithmic
deviation $\text{ln}(\varepsilon_{I_{c}})$ for $\Lambda/\omega_{m}=0.03$,
$0.043$ and $0.05$ {[}indicated by red dash lines in Fig. \ref{fig_bifurcation}
(a){]}, which correspond to the regular periodic oscillation regime,
the transition point and the period-doubling regime, respectively.
For $\Lambda/\omega_{m}=0.03$, the classical mean values of the cavity
and mechanical modes finally converge to a regular limit cycle for
the evolution time less than $30\tau$, and the cavity intensity $I_{c}$
oscillates with the period of $\tau=2\pi/\text{\ensuremath{\Omega}}$,
see Fig. \ref{fig_003}. In addition, the calculated exponential variation
of $\varepsilon_{I_{c}}$ infinitely approaches to zero as time evolves,
manifesting that all the nearby points of the initial cavity intensity
$I_{c}$ in phase space will finally oscillate in the identical frequency
$\Omega$. For $\Lambda/\omega_{m}=0.05$, the phase-space trajectory
of the cavity and mechanical modes and time-dependence of cavity intensity
then demonstrates the period-doubling motion. Remarkably, the dynamical
behavior in this regime can also be observed for a short evolution
time $\sim30\tau$. The fast decrease of $\text{ln}(\varepsilon_{I_{c}})$
over time also signifies that the system can evolve onto its stable
orbit rapidly. 

While for $\Lambda/\omega_{m}=0.043$, in the vicinity of transition
point, the system first experiences a transient period-doubling motion
{[}see the inset of Fig. \ref{fig_0043}(c){]}, and then displays
a sinusoidal-like oscillation after a long evolution time $\sim300\tau$,
as indicated by the limit-cycle in phase space {[}see Fig. \ref{fig_0043}(a)-(b){]}.
The translational behavior can be further understood by the slow-evolution
property of $\text{ln}(\varepsilon_{I_{c}})$ shown in Fig. \ref{fig_0043}(d)
by comparing with that for $\Lambda/\omega_{m}=0.03$ and $\Lambda/\omega_{m}=0.05$. 

\begin{figure}
\includegraphics[width=0.85\columnwidth]{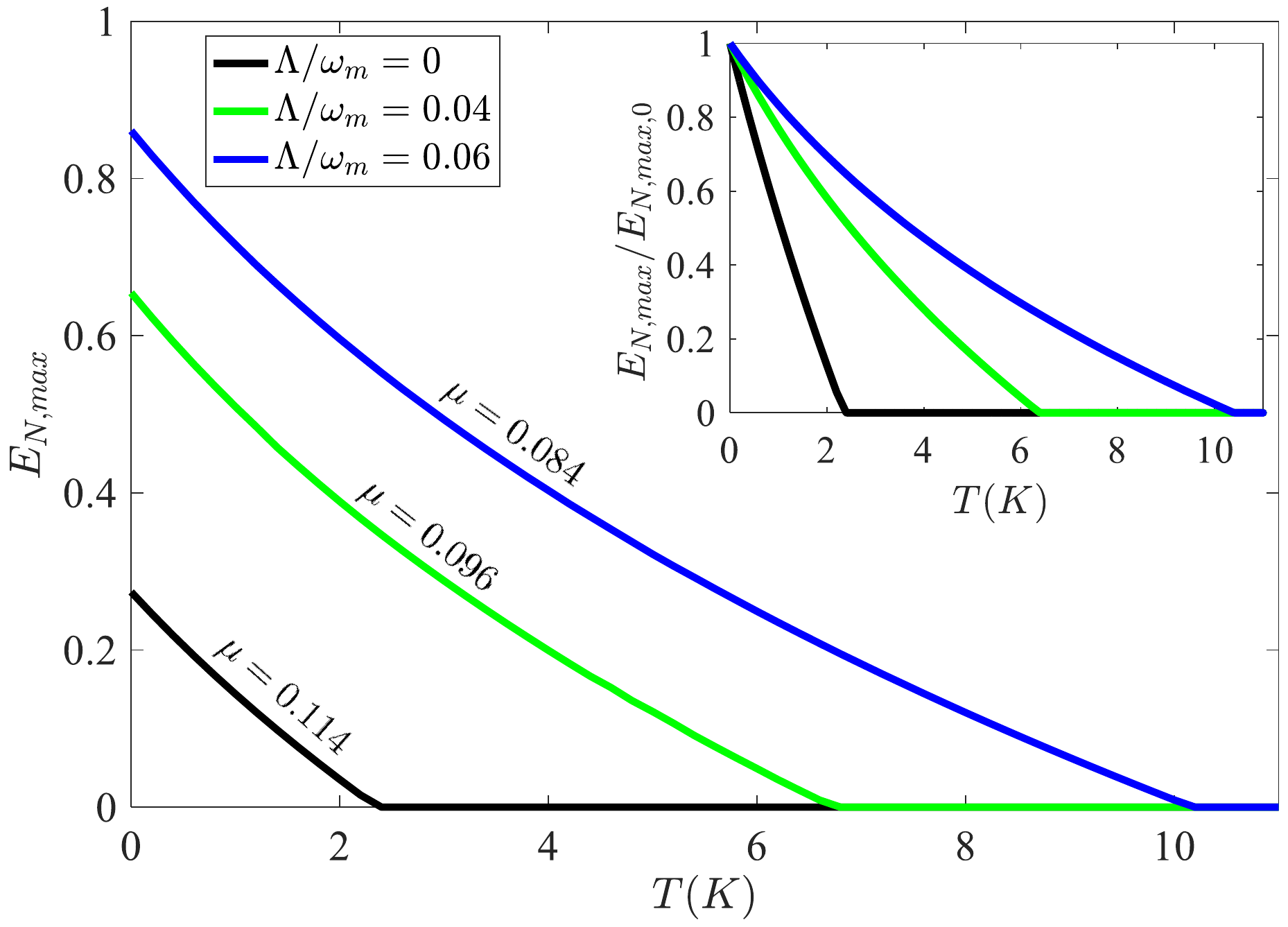}

\caption{\label{EN_T}(Color online) $E_{N,max}$ as a function of the thermal
temperature $T$ for $\Lambda/\omega_{m}=0$ (the black lower line), $0.04$ (the green middle line), and $0.06$ (the blue upper line). $\mu$ is the entanglement to temperature ratio
and the inset shows the renormalized $E_{N,max}$. Other parameter
are the same as in Fig. \ref{fig_bifurcation}.}
\end{figure}

In correspondence to the classical dynamics, the light-mechanical
pseudo-entanglement $E_{p}\equiv-\text{ln}2\eta$ for the parametric
gains $\Lambda/\omega_{m}=0.03$, $0.043$, $0.05$ exhibits a similar
dynamical behavior, as shown in Fig. \ref{EN_three}. For $\Lambda/\omega_{m}=0.03$
and $\Lambda/\omega_{m}=0.05$, the dynamical entanglement quickly
become stable in a short time, and thus, the time-dependence of $E_{p}$
for the time intervals {[}$30\tau$, $40\tau${]} and {[}$290\tau$,
$300\tau${]} are the same. While for $\Lambda/\omega_{m}=0.043$,
we again observe the transient and transitional dynamical behavior
in $E_{p}$. These are other quantum manifestations of the classical
dynamics. Moreover, it should be noted that the pseudo-entanglement
$E_{p}$ less than zero actually denotes non-existence of light-mechanical
entanglement, therefore, the system exhibits sudden death and rebirth
of light-mechanical entanglement over evolutional time.

Finally, the experimental observation of the above int\textcolor{black}{eresting
behavior can be realized with the set of parameters \cite{Ma_pra2014,Ying_pra2014,Tho_Nat2008}:
$L=25$ mm, $F=1.4\times10^{4}$ , $\omega_{m}=2\pi\times1$ MHz,
$Q=10^{6}$, $m=150$ ng and the driving laser power $P=21.9$ mW
and wavelength $\lambda=1064$ nm. In this case, the cavity decay
rate is given by $\kappa/2\pi=c/(4FL)\sim0.21$ MHz and the effective
optomechanical coupling strength for $\Lambda=0$ is around $g_{0}/\omega_{m}=0.48$.
Thus, the optomechanical cavity is in the resolved-sideband regime
as discussed previously, and the system stays far away from the rotating
wave approximation regime, where the off-resonant down conversion
process leads to the nonvanishing optomechanical entanglement at $T=0$
\cite{Genes_PRA2008}. Moreover, by considering a finite environmental
temperature $T$, we find that the light-mechanical entanglement can
be greatly enhanced in the presence of OPA, as shown in Fig. \ref{EN_T}.
In particular, the entanglement to temperature ratio $\mu\equiv E_{N,max}(T=0)/T_{E_{N,max}=0}$
as well as the renormalized entanglement $E_{N,max}(T)/E_{N,max}(T=0)$
show that the system is more robust against thermal noise while the
OPA p}umping strength increases, and thus, the period-doubling motion
(e.g. $\Lambda/\omega_{m}=0.06$) is less sensitive to environmental
temperature compared with regular periodic motion (e.g. $\Lambda/\omega_{m}=0.04$).

\section{CONCLUSION}

In summary, we have studied the classical nonlinear dynamics and the
light-mechanical entanglement dynamics in a cavity optomechanical
system involving an OPA under two different parameter regimes. In
one regime, the classical dynamical behavior (i.e. the regular periodic
oscillation and the period-doubling motion) can be manifested in the
optomechanical entanglement by special modulation of the OPA pumping
with the same parameters. For the other parameter condition that also
allows to generate mechanical squeezing, we can examine the normal
mode splitting effect in optomechanical entanglement, which finds
no classical correspondence. The dynamical behavior is also related
to the robustness of the system to the environmental temperature,
as evidenced by the death and revival of the optomechanical entanglement.
The standard optomechanical system with the assistance of OPA therefore
provides a nice architecture for studying and exploring the manifestation
of classical nonlinear dynamics in quantum entanglement. 
\begin{acknowledgments}
L.-T.S., Z.-B.Y., H.W., Y.L., and S.-B.Z are supported by the National
Natural Science Foundation of China under Grants No. 11774058, No.
11674060, No. 11874114, No. 11875108, No. 11705030, and No. 11774024,
the Natural Science Foundation of Fujian Province under Grant No.
2017J01401, and the Qishan fellowship of Fuzhou University. 
\end{acknowledgments}

\end{document}